\documentclass[3p,twocolumn]{elsarticle}

\usepackage{epsfig,epsf}
\usepackage{amsmath}
\usepackage{amsthm}
\usepackage{amsfonts}
\usepackage{amssymb}
\usepackage{dsfont}

\usepackage{psfrag}
\usepackage{slashed}

\def\OMIT#1{}

\newcommand \vev [1] {\langle{#1}\rangle}

\begin{document}

\begin{frontmatter}

\title{ Two-loop renormalization of three-quark operators in QCD}

\author[sk]{Susanne Kr\"ankl}
\ead{susanne.kraenkl@physik.uni-regensburg.de}
\author[sk,am1]{Alexander Manashov}
\ead{alexander.manashov@physik.uni-regensburg.de}

\address[sk]{Institut f\"ur Theoretisch Physik, University of Regensburg, D-93040 Regensburg,
Germany}

\address[am1]{ Department of Theoretical Physics,  St.-Petersburg State
University, 199034 St.-Petersburg, Russia}
\date{\today}

\begin{abstract}
Renormalization of composite three-quark operators in dimensional regularization is complicated by
the mixing of physical and unphysical (evanescent) operators.
This mixing must be taken into account in a consistent subtraction scheme.
In this work we propose a particular scheme that allows one to avoid the necessity of
additional finite renormalization and is convenient in QCD applications.
As an illustration, we calculate
the two-loop anomalous dimensions of local three-quark operators in this scheme.
 \end{abstract}

\end{frontmatter}

\section{Introduction}
The studies of hard exclusive processes including baryons will constitute a significant
part of the experimental program at the planned new facilities at JLAB (Newport News) and
FAIR (Darmstadt). QCD description of such reactions and also heavy hadron weak decays
with light baryons in the final state (LHCB, CERN) involves baryon
wave functions at small transverse separations, the so-called distribution amplitudes (DA).
The nucleon DAs
already have received considerable attention in the
literature,~see e.g.~Refs.~\cite{Lepage:1980fj,Chernyak:1984bm,Braun:2000kw,Braun:1999te}. In particular lattice
calculations can provide one with reliable estimates of the lowest moments of baryon
DAs which are defined by matrix elements of local three-quark
operators \cite{Braun:2008ur,Braun:2009jy}.

As usually, the QCD matrix elements depend on the scale and on the
renormalization scheme. The renormalization of baryon operators
in dimensional regularization beyond the leading logarithms involves some subtleties that, to our
knowledge,
have not been treated in the literature in a systematic way and the
purpose
of our letter is to fill this gap. Our study was fuelled in particular
by the need to calculate the NLO $\overline{\text{MS}}/\text{MOM}$ scheme conversion factors
in lattice studies (cf. \cite{Gracey:2011zn,Gracey:2011zg}) and the NLO extensions of light-cone sum rule
calculations of baryon form factors~\cite{Braun:2006hz} which have already started~\cite{PassekKumericki:2008sj}.

It is well known that for  a generic operator the widely used  $\overline{\text{MS}}$
prescription does not fix a renormalization scheme completely. Indeed, for a noninteger
dimension $d$ the Lorentz group $SO(1,d-1)$ becomes infinite dimensional. As a consequence
the tensor content of the $d-$dimensional theory is richer than that of a strictly four dimensional one.
In particular, it means that there exist operators in $d-$dimensions which have no
counterparts in four dimensions, e.g. totally antisymmetric tensors of rank $n>4$.
Such operators vanish in $d=4$ and are conventionally referred to as evanescent operators.
However, one cannot simply exclude evanescent operators from consideration since under
renormalization
they mix with the physical operators. The situation was thoroughly
analyzed by Dugan and Grinstein~\cite{Dugan:1990df}.
They have shown that one can always get rid of the mixing by a suitable finite
renormalization. A more detailed discussion  in~\cite{Herrlich:1994kh} shows that such finite renormalization
 is not unique.

We are interested in the renormalization of the local three-quark  operator
\begin{align}\label{Q}
\mathcal{Q}^{abc}_{\alpha\beta\gamma}=\epsilon^{ijk} q^{ia}_\alpha\,q^{jb}_\beta\,q^{kc}_\gamma\,.
\end{align}
Here $\alpha,\beta,\gamma$ are the spinor indices, while $ijk$ and $abc$ are the color and
flavor indices of the quark fields, respectively. The flavor structure of the operator
will be irrelevant for the further discussion, so that from now on we will suppress the
flavor indices. In typical applications one usually tries to get rid of multiple spinor
indices by contracting~(\ref{Q}) with suitable $\gamma-$matrices. An example is provided
by the so~--~called Ioffe current~\cite{Ioffe:1981kw}
\begin{equation}
\label{Ioffe}
  \eta_I(x) = \epsilon^{ijk}\left[u^i(x)C\gamma_\mu u^j(x)\right]\gamma_5\gamma^\mu d^k(x)\,,
\end{equation}
or the leading twist nucleon operator
\begin{align}
\label{eq:fn}
  \eta_N(x) = \epsilon^{ijk}\left[u^i(x)C \slashed{z}u^j(x)\right]\slashed{z} d^k(x)\,.
\end{align}
Here $z$ is a light-like auxiliary vector and $C$ is the charge conjugation matrix.
 The two-loop anomalous
dimension of the Ioffe current was calculated in~\cite{Pivovarov:1991nk}. Evanescent
operators do not contribute at this order to the current~(\ref{Ioffe}) due to special
cancelations however they should be taken into account for the current~(\ref{eq:fn}).

The renormalization of the currents~(\ref{Ioffe}),~(\ref{eq:fn}) is not of  academic
interest only~\footnote{The maximal
helicity three-quark operators in the \mbox{$\mathcal{N}=4$} SUSY were studied in
Refs.~\cite{Belitsky:2004sf,Belitsky:2005bu} in
AdS/CFT context.}. The sum rules involving these currents provide an important tool for quantitative study of
hadronic properties,
see e.g. Refs.~\cite{Chernyak:1984bm,Braun:2006hz,Ioffe:1981kw,King:1986wi}.
The matrix elements of the operator~(\ref{Q}) and the corresponding
interpolating currents satisfy at tree level certain symmetry relations which are derived
with help of Fierz transformations~\cite{Braun:2000kw}. Therefore it is preferable to use a renormalization
scheme which respects  these relations.

In our opinion the scheme developed in~~\cite{Dugan:1990df,Herrlich:1994kh,Pivovarov:1991nk}
is not best suited for renormalization of the three-quark operators.
We will suggest a different scheme and calculate the two-loop anomalous dimensions of the
operator~(\ref{Q}) in this scheme.

The paper is organized as follows: In Sect.~\ref{sect:scheme} we briefly review  the
renormalization approach taken by Dugan and Grinstein~\cite{Dugan:1990df}. Then we propose a
new renormalization scheme and discuss its properties. In Sect.~\ref{sect:two-loop}
we present the results of the calculation of anomalous dimensions of the operator~(\ref{Q})
in two loops. Conclusions are presented in Sect.~\ref{sect:summary}.

\section{Renormalization schemes}\label{sect:scheme}
Let us at first discuss the renormalization of the Ioffe current in the
scheme~\cite{Dugan:1990df}. It will be  more convenient to work with the current
$\bar\eta_1(x)=\gamma_5\eta_I(x)$ which definition does not involve the matrix $\gamma_5$.
Next, we will assume that the $d-$dimensional charge conjugation matrix $C$  satisfies the
defining relation $C\gamma_\mu C^{-1}=-\gamma^T_\mu$. Then one can easily  verify that
the operator $\bar \eta_1$ mixes under renormalization with the operators
\begin{align}\label{etan}
\bar \eta_n=\epsilon^{ijk}\left[u^iC\gamma^{(n)} u^j\right]\gamma^{(n)} d^k\,,
\end{align}
$n$ odd. Here  $\gamma^{(n)}\otimes \gamma^{(n)}
\equiv\gamma^{(n)}_{\mu_1\ldots\mu_n}\otimes\gamma^{(n)}_{\mu_1\ldots\mu_n}$
and $\gamma^{(n)}_{\mu_1\ldots\mu_n}$ is the antisymmetrized product of $n$ gamma
matrices,
$
\gamma^{(n)}_{\mu_1\ldots\mu_n}=\gamma_{[\mu_1}\ldots\gamma_{\mu_n]}\,.
$

The renormalized operators $\bar \eta_n$ take the
form~\footnote{In the $\overline{\text{MS}}$ scheme the renormalization matrix $Z$ is a series in $1/\epsilon$.}
\begin{align}\label{}
[\bar \eta_n]=\sum_{k=1,3\ldots}Z_{nk}\bar\eta_k\,.
\end{align}
Here and below the square brackets will denote  a renormalized operator.

In $d-$dimensions all operators~(\ref{etan})
are independent and satisfy the standard RG equation
\begin{align}\label{RG}
\Big([M\partial_M+\beta(\alpha_s)\partial_{\alpha_s}]\delta_{nk}+\gamma_{nk}\Big)[\bar \eta_k]=0\,.
\end{align}
Here $M$ is the renormalization scale, $\beta(\alpha_s)$ is the beta function and $\gamma_{nk}$
is the anomalous dimension matrix
\begin{align}\label{gamma}
\gamma=-M\left(\frac{d}{dM} \mathbb{Z}\right)\, \mathbb{Z}^{-1}\,, && \mathbb{Z}=Z Z_{q}^{-3}\,
\end{align}
with $Z_q$ being the quark field renormalization constant, $q_0=Z_q q$.

Generic correlation functions with insertion of  the operator $\bar \eta_n$, $n>3$, vanish at tree level in
$d=4$, thus these operators are evanescent ones. However, it does not hold anymore beyond
the tree level approximation. Moreover, even if the evanescent operators were zero at one scale
they would reappear at another scale due to mixing with physical operators.

It was shown in~\cite{Dugan:1990df}
that making a finite renormalization one can ensure vanishing of evanescent operators beyond
tree level at any scale. In other words this means that in an arbitrary scheme  the evanescent operators
 $\bar \eta_n$,
$n>3$
can be expressed in $d=4$ in terms of the first two, "physical", operators $\bar \eta_1$ and
$\bar\eta_3$~\footnote{
Strictly speaking one can choose any other two operators as the basic (physical) ones,
however in this case the expansion coefficients $a_{nk}$ will be singular in $\alpha_s$.
}
\begin{align}\label{n13}
[\bar \eta_n]=a_{n1}(\alpha_s)[\bar\eta_1]+a_{n3}(\alpha_s)[\bar\eta_3],&& n>3\,.
\end{align}
The above identity should be understood as an equality   between arbitrary correlation functions of
the operators on the l.h.s and on the r.h.s of this equation. (To avoid misunderstanding we remind that
taking the limit $d\to4$ one puts  $\gamma^{(n)}=0$
for  $n>4$ in the {\it renormalized} correlation functions.)

The  proof of Eq.~(\ref{n13})   given in~Ref.~\cite{Dugan:1990df} employs
combinatorics of the $R-$operation. It was done on the example of the four-fermion
operators but the argumentation is quite general.
Another way to verify Eq.~(\ref{n13}) is to note that in  some nonminimal schemes, e.g. the MOM$-$scheme,
the condition $[\bar\eta_n]=0$ for
$n>3$ holds automatically.  Operators in any two scheme are related by a {\it
finite} renormalization that implies~Eq.~(\ref{n13}).

Inserting~(\ref{n13})
into~(\ref{RG}) one derives the RG equation involving the physical operators $\bar \eta_{1,3}$
only
\begin{align}\label{RG-1}
\Big([M\partial_M+\beta(\alpha_s)\partial_{\alpha_s}]\delta_{nk}+\tilde \gamma_{nk}\Big)[\bar \eta_k]=0\,,
\end{align}
where the indices $n, k$ take the values $1,3$ and
\begin{align}\label{}
\tilde \gamma_{nk}= \gamma_{nk}+\sum_{p=5,7\ldots}\gamma_{np}a_{pk}\,.
\end{align}
Evidently one has to consider the evanescent operators in order to obtain
the correct result for the anomalous dimension matrix.

It is also clear that if the operators $\bar\eta_n$ enter the OPE  of certain currents
in $d-$dimensions, the four dimensional
OPE can be written in terms of the physical operators only with the coefficient
functions being appropriately modified.

\vskip 5mm
In this work we propose a different $\overline{\text{MS}}$ scheme for the renormalization of the
three-quark operator~(\ref{Q}) (see also the recent paper~\cite{Groote:2011my}).
 Our analysis follows closely the approach taken in~\cite{Vasiliev:1996rd}.
Instead of contracting the
operator~(\ref{Q}) with different $\gamma-$matrices we will consider the operator with
open spinor indices.
The divergent part of any diagram contributing to the correlator
\begin{align}\label{corr}
\vev{Q_{\alpha\beta\gamma}(0)\bar q_{\alpha'}(k)
\bar q_{\beta'}(p)\bar q_{\gamma'}(q)}
\end{align}
after subtracting the sub-divergences
can be cast into the form
\begin{align}\label{st}
\sum_{nmk}f_{nmk}(\epsilon)\, (\Gamma_{nmk})^{\alpha'\beta'\gamma'}_{\alpha\beta\gamma}\,,
\end{align}
where the functions $f_{nmk}(\epsilon)$ are series in $1/\epsilon$.
The gamma matrix
structures $\Gamma_{nmk}$ are defined by
\begin{align}\label{}
(\Gamma_{nmk})^{\alpha'\beta'\gamma'}_{\alpha\beta\gamma} =
\gamma^{(n)}_{\alpha\alpha'}\otimes\gamma^{(m)}_{\beta\beta'}\otimes\gamma^{(k)}_{\gamma\gamma'}\,,
\end{align}
where it is assumed that all Lorentz indices of gamma matrices ($\gamma^{(n)}_{\mu_1\ldots\mu_n}$) are
contracted between themselves~\footnote{There is only one nontrivial way to contract all indices.}.
We define the subtraction scheme  by  removing all singular  terms~(\ref{st}) from the
correlator~(\ref{corr}).
Thus the renormalized operator $\mathcal{O}_{\alpha\beta\gamma}$
takes the form
\begin{align}\label{}
[\mathcal{O}_{\alpha\beta\gamma}]=Z_{\alpha\beta\gamma}^{\alpha'\beta'\gamma'}\mathcal{O}_{\alpha'\beta'\gamma'},
\end{align}
where
\begin{align}\label{Z}
Z_{\alpha\beta\gamma}^{\alpha'\beta'\gamma'}=1+\sum_{nmk}
a_{nmk}(\epsilon)\,(\Gamma_{nmk})^{\alpha'\beta'\gamma'}_{\alpha\beta\gamma}
\end{align}
and
\begin{align}\label{}
a_{nmk}(\epsilon)=\sum_{p=1}^\infty{\epsilon^{-p}}\, a_{nmk}^{(p)}(\alpha_s)\,.
\end{align}
The RG equation for the operator $[\mathcal{O}_{\alpha\beta\gamma}]$ reads
\begin{align}\label{}
\Big(M\partial_M+\beta(\alpha_s)\partial_{\alpha_s}\Big)[\mathcal{O}_{\alpha\beta\gamma}]=
-\gamma^{\alpha'\beta'\gamma'}_{\alpha\beta\gamma}[\mathcal{O}_{\alpha'\beta'\gamma'}]\,,
\end{align}
where the anomalous dimension matrix $\gamma$ is given by Eq.~(\ref{gamma}).
Calculating the inverse matrix $Z^{-1}$ one has to carry out all gamma matrix algebra in
$d-$dimensions and this results in emergence of $\epsilon-$regular contributions in $Z^{-1}$.
Thus the matrix $Z^{-1}$ contains both singular and regular terms in $\epsilon$
\begin{align}\label{}
(Z^{-1})_{\alpha\beta\gamma}^{\alpha'\beta'\gamma'}=
1+\sum_{nmk}\tilde a_{nmk}(\epsilon)\, (\Gamma_{nmk})^{\alpha'\beta'\gamma'}_{\alpha\beta\gamma}\,,
\end{align}
with
$
\tilde a_{nmk}(\epsilon)=\sum_{p=-\infty}^\infty{\epsilon^{-p}}\, \tilde a_{nmk}^{(p)}(\alpha_s)\,.
$
This is different from the standard situation where $Z^{-1}$ is a series in $1/\epsilon$,
$Z^{-1}=\sum_k \tilde{a}_k (1/{\epsilon}^k)$ and there are no finite in $\epsilon$ terms.

The anomalous dimension matrix $ \gamma $ in $d-$dimensions takes the  form similar to~(\ref{Z})
\begin{align}\label{gamma1}
\gamma=\sum_{nmk}\gamma_{nmk}(\alpha_s,\epsilon)\, \Gamma_{nmk}\,,
\end{align}
(we omitted the spinor indices for brevity). The coefficients
$\gamma_{nmk}(\alpha_s,\epsilon)$ are regular functions of $\alpha_s$ and $\epsilon$.
In $d=4$ one can drop  all   $\Gamma-$matrices which  vanish in four
dimensions from the sum~(\ref{gamma1}), i.e. $n,m,k\leq 4$.

Let us see what happens with evanescent operators in this scheme. We define the
renormalized current $[\bar\eta_n]$ as follows
\begin{align}\label{retan}
[\bar \eta_n]=(C\gamma^{(n)})_{\alpha\beta}\otimes\gamma^{(n)}_{\gamma'\gamma}\,
[\epsilon^{ijk}u^{i}_\alpha u^{j}_\beta d^{k}_{\gamma}]\,.
\end{align}
Since the  $\gamma-$matrix structure is convoluted with the renormalized (finite) operator, one can
safely put $\gamma^{(n)}=0$ for $n\geq 5$ in $d=4$. Thus, the evanescent operators automatically vanish in $d=4$
in this scheme and we can treat the renormalized operator
$[\mathcal{O}_{\alpha\beta\gamma}]$ as a pure four dimensional object.
Since the $\gamma-$matrices in~Eq.~(\ref{retan}) are already four dimensional
one avoids  the problem of
defining  the matrix $\gamma_5$  in $d-$dimensions. Moreover, this scheme is obviously consistent with
the Fierz identities, as all contractions of the
renormalized operator with $\gamma-$matrices are done in four dimensions.

\section{Two loop analysis}\label{sect:two-loop}

 \begin{figure}[tb]
 \begin{center}{\includegraphics[width=0.25\textwidth,clip=true]{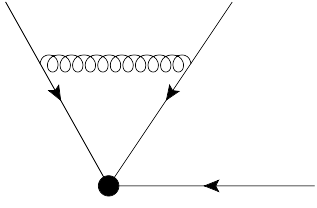}}
 \end{center}
 \caption{The one-loop correction to the operator $\mathcal{O}_{\alpha\beta\gamma}$.
 }
 \label{fig:one-loop}
 \end{figure}

In this section we present results of  the two-loop calculation of the anomalous dimension matrix
for the three-quark operator~(\ref{Q}). All calculations were done in  Feynman gauge.
The quark field renormalization constant $Z_q$ in this gauge reads~\cite{Egorian:1978zx}
\begin{align}\label{zq}
Z_q=&1-
\frac{a}{2\epsilon}C_F+
\frac{a^2}{2}\biggl[
\frac{1}{\epsilon^2}\left(\frac14 C_F^2+C_FC_A\right)
\notag\\
&+\frac1\epsilon C_F\left(
-\frac{17}{4}C_A+\frac34C_F+\frac12N_f
\right)
\biggr]\,.
\end{align}
Here $a=\alpha_s/(4\pi)$, $C_F=(N_c^2-1)/2N_c$, $C_A=N_c$
and $N_f$ is the number of flavors. The beta function for the coupling $a$
is
\begin{align}\label{beta}
\beta_a(a)=M\frac{da}{dM}=-2\epsilon a +2b_0 a^2+O(a^3)\,,
\end{align}
where $b_0=2N_f/3-11/3 N_c$.

The  expression for the
anomalous dimension matrix~(\ref{gamma}) takes the form
\begin{align}\label{gamma2}
\gamma=&
2a\biggl\{\epsilon
{\mathbb{Z}}^{(1)}
+a\Big(\epsilon\Big[2{\mathbb{Z}}^{(2)}- {\mathbb{Z}}^{(1)}{\mathbb{Z}}^{(1)}\Big]
\notag\\
&
\phantom{2a}-b_0{\mathbb{Z}}^{(1)}\Big)\biggl\}+O(a^3)\,,
\end{align}
where
\begin{align}
\mathbb{Z}=Z Z_{q}^{-3}=1+a\, \mathbb{Z}^{(1)}+a^2 \mathbb{Z}^{(2)}+O(a^3).
\end{align}
Calculating the one-loop diagram shown in Fig.~\ref{fig:one-loop} one gets the following
expression for the renormalization matrix $Z^{(1)}$  ($Z=1+\sum a^k Z^{(k)}$)
\begin{align}\label{}
Z^{(1)}=-\frac{1}{6\epsilon}\Big[\Gamma_{220}+\Gamma_{202}+\Gamma_{022}+12
\Big]\,,
\end{align}
where
\begin{align}\label{}
\Gamma_{220}=\gamma^{(2)}_{\mu\nu}\otimes \gamma^{(2)}_{\mu\nu}\otimes I
\end{align}
and similarly for the others. The one-loop anomalous dimension matrix~(\ref{gamma2})
takes the form
\begin{align}\label{}
\gamma=-\frac{a}{3}\Big[\Gamma_{220}+\Gamma_{202}+\Gamma_{022}\Big]\,.
\end{align}
In order to present the results of the two-loop calculations in a compact form we introduce
short--hand notations for combinations of $\gamma-$matrix structures which appear in
the calculations
\begin{align}\label{CC}
&\mathbb{C}_2=\,\Gamma_{220}+\Gamma_{202}+\Gamma_{022}\,,\notag\\
&\mathbb{C}_4=\,\Gamma_{440}+\Gamma_{404}+\Gamma_{044}\,,\notag\\
&\mathbb{C}_{42}=\Gamma_{422}+\Gamma_{242}+\Gamma_{224}\,,
\end{align}
where
\begin{align*}\label{}
\Gamma_{440}=&\gamma^{(4)}_{\mu_1\mu_2\mu_3\mu_4}\otimes \gamma^{(4)}_{\mu_1\mu_2\mu_3\mu_4}\otimes I\,,
\notag\\[2mm]
\Gamma_{422}=&\gamma^{(4)}_{\mu_1\mu_2\mu_3\mu_4}\otimes \gamma^{(2)}_{\mu_1\mu_2}\otimes
\gamma^{(2)}_{\mu_3\mu_4}\,.
\end{align*}

The contributions to the matrix $Z$ from  the two-loop diagrams
read
\begin{align}\label{2loop}
Z^{(2a)}=&-\frac29\left\{\frac1{\epsilon^2}\Big[\mathbb{C}_2+12\Big]
+\frac1{2\epsilon}\mathbb{C}_2\right\}\,,
\notag\\[2mm]
Z^{(2b)}=&\frac13\left\{\frac{5}{4}\left(\frac1{\epsilon^2}-\frac{17}{30\epsilon}\right)-
\frac{N_f}{6}\left( \frac{1}{\epsilon^2}-\frac1{6\epsilon}\right)\right\}\mathbb{C}_2\,,\notag\\
Z^{(2c)}=&-\frac1{36}\left\{\frac1{\epsilon^2}\Big[\mathbb{C}_2
+12\Big]+\frac5{2\epsilon}\mathbb{C}_2\right\}\,,
\notag\\[2mm]
Z^{(2d)}=&\frac3{4}\left\{\frac1{\epsilon^2}\Big[\mathbb{C}_2
+12\Big]+\frac1{\epsilon}\left(\frac16\mathbb{C}_2-4\right)\right\}\,,
\notag\\[2mm]
Z^{(2e)}=&\frac1{\epsilon}\frac{5}{144}
\Big\{
\mathbb{C}_4-4\mathbb{C}_2-24
\Big\}\,,
\notag\\[2mm]
Z^{(2f)}=&\frac1{36}\biggl\{
\frac{1}{\epsilon^2}\left(\frac12 \mathbb{C}_4+8\mathbb{C}_2+60\right)
\notag\\
&\phantom{36}-\frac1\epsilon\left(\frac14\mathbb{C}_4+12\mathbb{C}_2+120\right)
\biggr\}\notag \, ,\\[2mm]
Z^{(2g)}=&\frac1{36}\biggl\{
\frac{1}{\epsilon^2}\left( \mathbb{C}_{42}+6\mathbb{C}_2+48\right)
\notag\\
&\phantom{36}
-\frac1\epsilon\left(\frac12\mathbb{C}_{42}+\mathbb{C}_2\right)
\biggr\}\,.
\end{align}
Here the factor  $Z^{(2a)}$ $(Z^{(2b)})$ comes from the quark (gluon) self-energy corrections
to the one-loop diagram shown in Fig.~\ref{fig:one-loop}. Similarly, $Z^{(2c)}$ and  $Z^{(2d)}$
correspond to the QED~--~like and QCD~--~like vertex corrections, respectively.
$Z^{(2e)}$ and $Z^{(2f)}$ result from to the diagrams with intersecting and
non~--~intersecting gluon lines, respectively.
Finally, $Z^{(2g)}$ takes into account
contributions from the diagrams with all three quarks interacting~\footnote{The diagram where all quark lines
 connected via triple gluon vertex vanishes because of the color structure.}.

The calculations are straightforward so we will not present them. Some
useful identities for the $d$--dimensional $\gamma$--matrices can be found in
Refs.~\cite{Dugan:1990df,Vasiliev:1995qj}. We also checked that our results are consistent
with the two-loop calculations of Pivovarov and Surguladze~\cite{Pivovarov:1991nk}.

Substituting~(\ref{2loop}) and~(\ref{zq}) into (\ref{gamma2}) and taking into account the identity
\begin{align*}
\mathbb{C}_{42}=\frac12\mathbb{C}_2^2-\frac12\mathbb{C}_4-2(d-3)\mathbb{C}_2-3d(d-1)
\end{align*}
one obtains for the anomalous dimension matrix in~$d=4$
\begin{align}\label{gamma12}
\gamma=&-\frac{a}{3}\mathbb{C}_2-
\left(\frac{a}{6}\right)^2\biggl\{\mathbb{C}_2^2-{5}\mathbb{C}_4
-2(b_0-36)\mathbb{C}_2\biggr\}\notag\\[1mm]
&-6a^2(b_0-1)+O(a^3)\,.
\end{align}
To find the eigenvalues and eigenvectors of the matrix $\gamma$ let us note that
the matrices $\mathbb{C}_2$ and $\mathbb{C}_4$ entering~(\ref{gamma12}) are Lorentz invariant operators.
An operator $\mathcal{O}^{(j\bar j)}$ transforming according to the
irreducible representation of the Lorentz group which is labeled by two  spins~$(j,\bar j)$
diagonalizes the matrix $\mathbb{C}_2$, i.e.
$\mathbb{C}_2\,\mathcal{O}^{(j\bar j)}={c}_2^{(j\bar j)}\,\mathcal{O}^{(j\bar j)}$.
The corresponding eigenvalues are
\begin{align}\label{}
c_2^{(1/2,0)}=-c_2^{(3/2,0)}=&-3\,c_2^{(1,1/2)}=12\,,
\end{align}
and ${c}_2^{(j\bar j)}={c}_2^{(\bar j j)}$. The matrix $\mathbb{C}_4$ in $d=4$ has the
form
\begin{align*}
\mathbb{C}_4=
24(\gamma_5\otimes\gamma_5\otimes~I+\gamma_5\otimes I\otimes \gamma_5+I\otimes\gamma_5\otimes
\gamma_5)\,.
\end{align*}
Its eigenvalues  depend on  the chirality  of  the quark fields
entering an eigenoperator $\mathcal{O}^{(j\bar j)}$. Namely, $c_4^{+}=3\cdot 24$ if all quark fields have the
same chirality, and $c_4^{-}=-24$ otherwise.

Thus an eigenoperator carries three quantum
numbers~\footnote{The anomalous dimensions are insensitive to the flavor structure of the operator so that
we disregard it.}: Lorentz spins $(j,\bar j)$ and "chirality
$\pm$".

In the following we present four operators featuring different quantum numbers.
These are the standard operators which appear in the studies of the nucleon matrix
elements~\cite{Chernyak:1984bm,Braun:2000kw,Braun:2006hz,Ioffe:1981kw,King:1986wi}.
Two of these operators are of twist three and can be written in the form~\cite{Braun:1998id}
\begin{align}\label{tw3}
\mathcal{O}^{(\frac32,0)}_{+}=&\epsilon^{ijk} \slashed{z}q_L^i\,\slashed{z}q_L^j\,\slashed{z}q_L^k\,,
\notag\\
\mathcal{O}^{(1,\frac12)}_{-}=&\epsilon^{ijk} \slashed{z}q_L^i\,\slashed{z}q_L^j\,\slashed{z}q_R^k\,,
\end{align}
where $q_{L(R)}=\dfrac12(1\mp\gamma_5)q$ are left(right)-handed spinors and $z$ is a auxiliary light-like vector,
$z^2=0$.
The other two operators have twist four and can be written as follows~\cite{Pivovarov:1991nk}
\begin{align}\label{tw4}
\mathcal{O}^{(\frac12,0)}_{+}=&\epsilon^{ijk} (q_L^{iT}\,C\,q_L^j)\,q_L^k\,,
\notag\\
\mathcal{O}^{(\frac12,0)}_{-}=&\epsilon^{ijk} (q_L^{iT}\,C\,q_L^j)\,q_R^k\,.
\end{align}
Converting the operators~(\ref{tw3}),~(\ref{tw4}) into the standard notation one finds that
$\mathcal{O}^{(\frac12,0)}_{-}$ and $\mathcal{O}^{(1,\frac12)}_{-}$ correspond to the Ioffe current ${\eta}_I$,
Eq.~(\ref{Ioffe}), and the leading twist nucleon  current ${\eta}_N$, Eq.~(\ref{eq:fn}), respectively.
The operator $\mathcal{O}^{(\frac12,0)}_{+}$
is related to the Dosch current~\cite{Chung:1981cc}
\begin{align}
{\eta}_D(x)={\epsilon}^{ijk}\left[u^i(x)C{\sigma}_{{\mu}{\nu}}u^j(x)\right]\,{\gamma}_5{\sigma}^{{\mu}{\nu}}d^k(x)
\end{align}
whereas $\mathcal{O}^{(\frac32,0)}_{+}$ belongs to the baryon decuplet current defined in~\cite{Braun:1999te}.

For the anomalous dimensions of the  operators~(\ref{tw3}),~(\ref{tw4}) we obtain
\begin{align}\label{}
\gamma_+^{(\frac32,0)}=&\left(\frac{\alpha_s}\pi\right)+
\left(\frac{\alpha_s}\pi\right)^2\left(\frac{9}{4}-\frac5{12}b_0\right)\,,
\notag\\
\gamma_-^{(1,\frac12)}=&\frac13\left(\frac{\alpha_s}\pi\right)+
\left(\frac{\alpha_s}\pi\right)^2\left(\frac{23}{36}-\frac7{18}b_0\right)\,,
\notag\\
\gamma_+^{(\frac12,0)}=&-\left(\frac{\alpha_s}\pi\right)+
\left(\frac{\alpha_s}\pi\right)^2\left(-\frac{3}{4}-\frac1{3}b_0\right)\,,
\notag\\
\gamma_-^{(\frac12,0)}=&-\left(\frac{\alpha_s}\pi\right)+
\left(\frac{\alpha_s}\pi\right)^2\left(-\frac{19}{12}-\frac1{3}b_0\right)\,.
\end{align}

The anomalous dimensions of the operator subset~(\ref{tw4}) were calculated in~\cite{Pivovarov:1991nk}
in another scheme. The operators in these two different schemes are related to each other by a finite
renormalization
\begin{align}\label{}
\mathcal{O}^{(\frac12,0)}_{\pm}=\left(1-\frac{7}{3}\,a+O\bigl(a^2\bigr)\right)\Big(\mathcal{O}^{(\frac12,0)}_{\pm}\Big)^{\text{\cite{Pivovarov:1991nk}}}\,.
\end{align}
It can be checked that this factor accounts for the mismatch of the anomalous dimensions in the two schemes.

\section{Summary}\label{sect:summary}
We  proposed a simple scheme for the renormalization of local three-quark operators in
dimensional regularization. An attractive property of this scheme is the guaranteed
vanishing of the evanescent operators in $d=4$ dimensions so that one can work with
physical (four dimensional) operators only. The renormalization procedure maintains
explicitly all identities for operators based on Fierz transformations.

We have calculated the anomalous dimension matrix at two-loop order and found its
eigenvalues. Our results for the twist four operators agree with the known in literature.
The results for the leading twist three operators are new. In particular, the anomalous dimension
$\gamma_-^{(1,\frac12)}$ determines the scale dependence of the nucleon wave function at
origin,~$f_N$.

\section*{Acknowledgments}
We are grateful to I.V.~Anikin, V.M.~Braun and A.A.~Pivovarov for the valuable discussions. This work was
supported by DFG (grant 9209506, A.M.) and RFFI (grant
09-01-93108, A.M.).

\end{document}